\documentclass[times,twocolumn,twocolappendix]{aastex631}
\usepackage{amsmath,amssymb}	
\usepackage{subfigure,wrapfig}
\usepackage{xspace}
\usepackage{xcolor}

\newcommand{\rstar}{\ensuremath{R_{\star}}}
\newcommand{\bhac}{\texttt{BHAC}\xspace}

\begin{document}

\title{Three-dimensional GRMHD simulations of neutron star jets}

\author[0000-0002-4764-6189]{Pushpita Das}
\correspondingauthor{Pushpita Das}
\email{p.das2@uva.nl}
\affiliation{Anton Pannekoek Institute for Astronomy, University of Amsterdam\\
Science Park 904, 1098 XH, the Netherlands}

\author[0000-0002-4584-2557]{Oliver Porth}
\affiliation{Anton Pannekoek Institute for Astronomy, University of Amsterdam\\
Science Park 904, 1098 XH, the Netherlands}



\begin{abstract}
Neutron stars and black holes in X-ray binaries are observed to host strong collimated jets in the hard spectral state. Numerical simulations can act as a valuable tool in understanding the mechanisms behind jet formation and its properties. Although there have been significant efforts in understanding black-hole jets from general-relativistic magnetohydrodynamic (GRMHD) simulations in the past years, neutron star jets, however, still remain poorly explored. We present the results from three-dimensional (3D) GRMHD simulations of accreting neutron stars with oblique magnetospheres for the very first time. The jets in our simulations are produced due to the anchored magnetic field of the rotating star in analogy with the Blandford-Znajek process. We find that for accreting stars, the star-disk magnetic field interaction plays a significant role and as a result, the jet power becomes directly proportional to ${\Phi^2}_{\rm jet}$, where $\Phi_{\rm jet}$ is the open flux in the jet. The jet power decreases with increasing stellar magnetic inclination and finally for an orthogonal magnetosphere, it reduces by a factor of $\simeq 2.95$ compared to the aligned case. We also find that in the strong propeller regime, with a highly oblique magnetosphere, the disk-induced collimation of the open stellar flux preserves parts of the striped wind resulting in a striped jet. 
\end{abstract}



\section{Introduction} \label{sec:intro}

One of the key features of disk-fed accreting systems is the outflow of material either in the form of collimated jets or conical winds. This occurs in systems hosting various central objects ranging from neutron stars and black-holes in X-ray binaries (XRBs), black holes in active galactic nuclei (AGN) to young stellar objects (YSOs), and white dwarfs. Several mechanisms have been suggested to explain the formation of outflows from the accretion disk \citep[e.g.][]{BegelmanMcKee1983, Icke1980, VitelloShlosman1988} which can all contribute to a larger or lesser extent for a given system \citep{Proga2007,PudritzOuyed2007}. The mechanisms of relativistic jet formation from black holes and neutron stars are most commonly ascribed to the theoretical models suggested by \cite{BlandfordZnajek1977} and \cite{BlandfordPayne1982}. The Blandford-Payne (BP) process suggests that the accretion disk threaded by large poloidal magnetic fields can power astrophysical jets. The Blandford-Znajek (BZ) mechanism on the other hand relies on the frame-dragging effect of a rotating black hole which, if threaded by poloidal magnetic fields, launches a Poynting flux driven outflow along the field lines.  For compact stars, a third option exists in which the field lines are rooted in the rotating star, and a jet similar to the BZ process is launched from the magnetically dominated open field lines near the polar cap \citep{Parfrey2016,ParfreyTchek2017,Das2022}.   

Jets in X-ray binaries are most commonly studied in the radio band. The radio emission originates close to the central object and is typically associated with non-thermal emission from electrons in the jet \citep{Falcke2000}. The evolution of the jet is strongly coupled to the nature of the accretion flow in the vicinity of the compact object. In  black-hole X-ray binaries (BHXRBs) a steady radio jet is often observed in the hard X-ray state, during transition to the soft state the jet grows transitional and more powerful, and is  finally suppressed during the soft state \citep{Fender2004}.  Similar to BHXRBs a strong radio jet is also observed during hard states in neutron star X-ray binaries (NSXRBs). However, as the system evolves towards soft X-ray states, contrary to black-holes, some show signs of jet suppression \citep{Miller-Jones2010} and some do not \citep{MigliariFender2004}.   Along with the differences in jet behaviour with spectral states, for similar X-ray luminosity, neutron star jets are also observed to be radio-fainter by a factor of $\sim$ 10 \citep{GalloDegenaar2018,vandenEijnden2021} compared to their black hole counterparts and it has been argued that this difference cannot be explained by the additional X-ray emission from the surface \citep{FenderKuulkers2001,MigliariFender2006,Tudor2017}.  

There have been substantial studies on the radio vs. X-ray correlation in X-ray binaries \citep{HannikainenHunsteadEtAl1998,MerloniHeinz2003,Fender2009,GalloMiller-JonesEtAl2014,Carotenuto2021,GalloDegenaar2018,vandenEijnden2021}. 
Besides the systematic radio-quietness of neutron star systems, these studies have shown that NSXRBs are more ``varied'' than BHXRBs.  For example, individual neutron star systems that are otherwise very similar (in terms of $\dot{M}$, mass, spin, and magnetic field strength) show over an order of magnitude difference in radio power \citep[e.g.][]{RussellDegenaarEtAl2018, TetarenkoBahramianEtAl2018}.   Furthermore, several sources display large uncorrelated jumps in either radio- or X-ray power on timescales of days \citep[e.g.][]{TudorMiller-JonesEtAl2017,vandenEijndenDegenaarEtAl2020}.  
The differences in black hole and neutron star jet properties are often attributed to a different disk-jet coupling mechanism, however, no consensus has been reached yet.  

Numerical simulations of star-disk systems can be used to study jet formation and disk-jet coupling in NSXRBs. Stellar outflows and star-disk magnetic interaction have been extensively studied with non-relativistic MHD simulations \citep{MattPudritz2005a,ZanniFerreira2013,LovelaceRomanova2014, LiiRomanova2014, RomanovaBlinova2018,TakasaoTomida2022} and relativistic simulations \citep{Parfrey2016}. Axisymmetric GRMHD simulations of accreting neutron stars have been recently performed by \cite{ParfreyTchek2017,Das2022,SercanRezzolla2022}.  Due to their two-dimensional nature, the previous simulations were restricted to aligned configurations of the (quadru-) dipolar stellar field with the disk magnetic initial field configuration.  In this paper, we show results from the first 3D GRMHD simulations of accreting neutron stars and investigate the role of stellar magnetic inclination in jet formation.  
\section{Numerical Setup}\label{sec:setup}
We solve the following ideal MHD equations in a general relativistic framework using The Black Hole Accretion Code ($\bhac$) \citep{Porth2017, Olivares2019},
\begin{align}
    \nabla_{\mu}(\rho u^{\mu}) &= 0 \\
    \nabla_{\mu} T^{\mu\nu} &= 0 \\
    \nabla_{\mu} ^{\star}F^{\mu\nu} &= 0 \, .
\end{align}
Here, $T^{\mu\nu}$ is the energy-momentum tensor of an ideal magneto-fluid; $^{\star}F^{\mu\nu}$ is the dual of the Faraday tensor $F^{\mu\nu}$ and $\rho$,  $u^{\mu}$, are rest-mass density and fluid 4-velocity respectively. We perform the simulations in Cartesian Schwarzschild coordinates corotating at the stellar angular frequency ($\Omega_{\rm star}$). The metric components for our tailor-made coordinate system are given in Appendix \ref{appendix1}.

We initialize our simulation domain with a standard Fishbone-Moncrief torus \citep{Fishbone1976}, with the inner edge and density maximum located at $r_{\rm in} = 45 r_g$ ($r_g = GM/c^2$) and $r_{\rm max} = 65 r_g$ respectively. The torus is weakly magnetized with poloidal loops defined as $\rm A_{\phi} \propto max(\rho/\rho_{\rm max} - 0.1, 0)$ such that $2 p_{ \rm max}/b^2_{ \rm  max} \approx 110$ (where the subscript max indicates the individual maxima of the quantities as customary in black hole torus setups, \citealt{Porth2019}). Outside the torus, the magnetosphere is initialized with a stationary dipolar vector potential in Schwarzschild space-time following \citet{Wasserman1983}. 

Inside the magnetosphere ($r < r_{lc} =$ $ c/\Omega$), the initial pressure and density are set such that magnetization ($\sigma_t = b^2/\rho$) and plasma beta ($\beta_t = 2 p/b^2$) are 70 and 0.014 respectively. Both $\sigma$ and $\beta$ transition smoothly to follow an $r^{-6}$ profile outside the light-cylinder radius. We perturb the pressure with 8$\%$ white noise to excite the magnetorotational instability (MRI) inside the torus. The development of MRI results in angular momentum transport in the disk \cite{Balbus1991} and leads to accretion onto the neutron star.

In order to maintain force-free like conditions in the magnetosphere, we modify the numerical solution to drive pressure, density, and $v_{\parallel}$ to target values ($\rho_t = b^2/\sigma_t$, $p_t = \beta_t b^2/2$ and $v_{\|,t} = 0$,  see \cite{Das2022} for more details). In order to differentiate the accreting region from the magnetosphere, we introduce a tracer-fluid ($\tau$). The accretion disk is initialized with $\tau = 1$ and the rest of the domain is initialized with as $\tau = 0$. The tracer is passively advected with the flow by solving the extra equation $\nabla_{\mu}({\tau\rho u^{\mu}}) = 0$. Inside the magnetosphere (regions with $\tau = 0$), we drive the solution to the target values as mentioned above, for $0 < \tau < 1$ we mix the force-free and MHD solutions by interpolating between the two solutions. Finally, for $\tau = 1$, the solution follows that of the unmodified ideal GRMHD equations.  We refer the reader to \cite{ParfreyTchek2017} for further discussion of this approach to couple the force-free and ideal MHD regimes. For numerical stability, along with driving the solutions to target values, we also maintain global floor and ceiling values for $\beta$ and $\sigma$ such that at any point in time $\beta > 0.1\beta_t$ and $\sigma < 10 \sigma_t$.

Since the star is part of the computational grid, the stellar boundary conditions require special attention.  Inside the stellar radius, we set pressure and density to initial values such that $\beta = 0.014$ and $\sigma = 70$, and the velocity components ($v^x, v^y, v^z$) are overwritten such that the star maintains a solid body rotation with an angular frequency $\Omega$. Thereby, matter that reaches the stellar surface is effectively removed from the simulation, emulating the outflow boundary conditions used in spherical setups \citep{ParfreyTchek2017, Das2022}.
We update the magnetic field using the constrained transport algorithm of \texttt{BHAC} \citep{Olivares2019}. To fix the stellar magnetic field to the initial oblique dipole, within $r/R_\star<0.95$ we enforce that the co-rotating electric fields smoothly transition to zero using a sigmoid function. This avoids any kinks of the magnetic field-lines at the stellar boundary. The induction equation is solved in the entire domain, ensuring that no magnetic monopoles are created throughout the evolution. We tested that the method of flux freezing preserves the initial configuration with sufficient accuracy.  In particular, the relative difference in the magnetic flux (at r = $\rstar$) between the initial (t = 0$r_g/c$) and final state (t = 25000$r_g/c$) is $\simeq 5 \times 10^{-3}$. In order to test our stellar boundary conditions and driving criterion, we have also performed flat-spacetime isolated neutron star simulations for the aligned ($\chi_{\rm star} = 0^{\circ}$) and the orthogonal case ($\chi_{\rm star} = 90^{\circ}$). 
We find that the Poynting flux for these two cases closely follow the scaling obtained in the force-free simulations by \citep{Spitkovsky2006} with, in our case, $k_1 \simeq 1.1$ and $k_2 \simeq 0.9$.

For numerical integration, we use the total variation diminishing Lax-Friedrich scheme for fluxes along with Piecewise Parabolic reconstruction scheme and second order modified Euler time-stepper. Our simulation domain extends from $-400 \rm r_g$ to $400 \rm r_g$ in x, y, and z.\footnote{We extend the domain along the z-axis such that z $\in [-800,800] \rm r_g$ for two runs, one with $\Omega = 0.05$, $\mu = 30$, $\chi_{\rm star} = 60^{\circ}$ (propeller case) and the other $\Omega = 0.03$, $\mu = 20$, $\chi_{\rm star} = 30^{\circ}$ to understand the jet collimation.}. The stellar radius for all the runs is set to $\rstar = 4\rm r_g$. The base resolution of the domain is $128 \times 128 \times 128$ cells with 7 levels of AMR. This allows us to resolve dynamically interesting regions with an effective resolution of $8192^3$ cells; in particular, the stellar radius is resolved by over 40 cells. Unless stated otherwise, throughout this letter we adopt geometric units with $G = c = 1$. 

\section{Results}

\begin{figure*}
   \centering
     \includegraphics[width=18.3cm, angle=0]{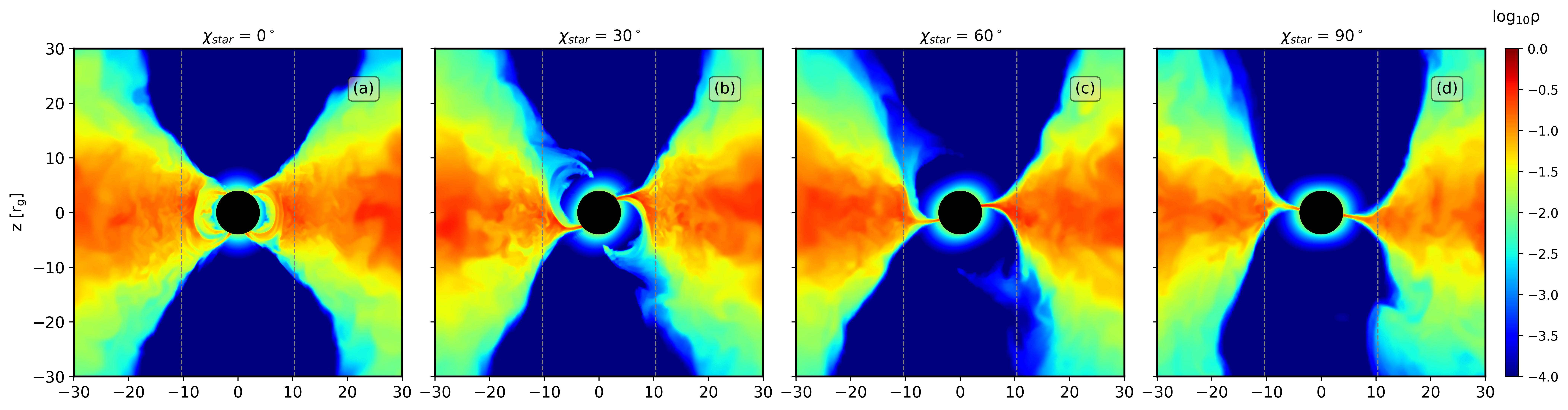}
     \includegraphics[width=18.3cm, angle=0]{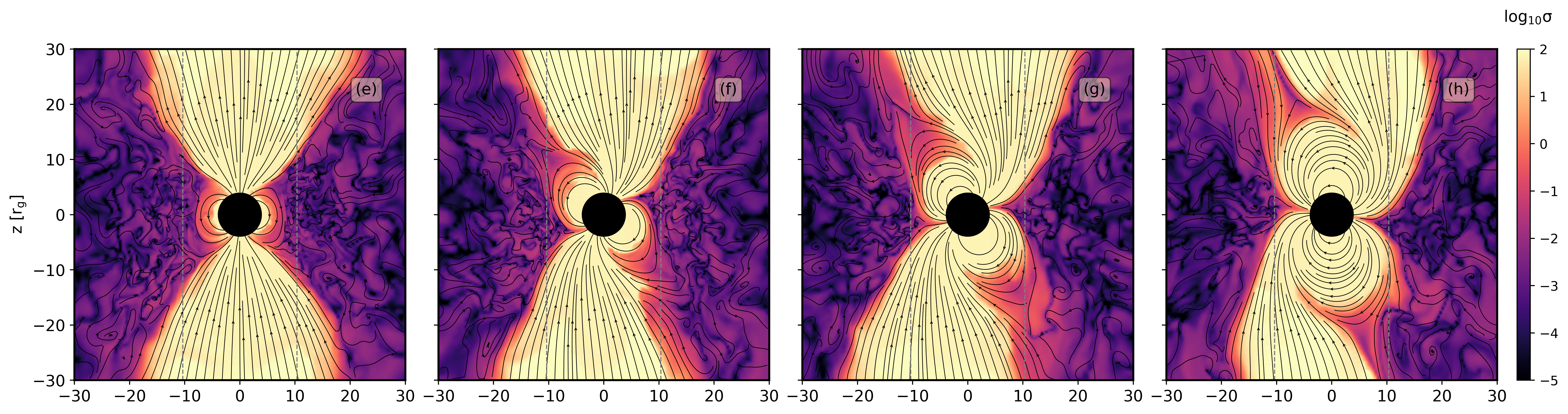}
     \includegraphics[width=18.5cm, angle=0]{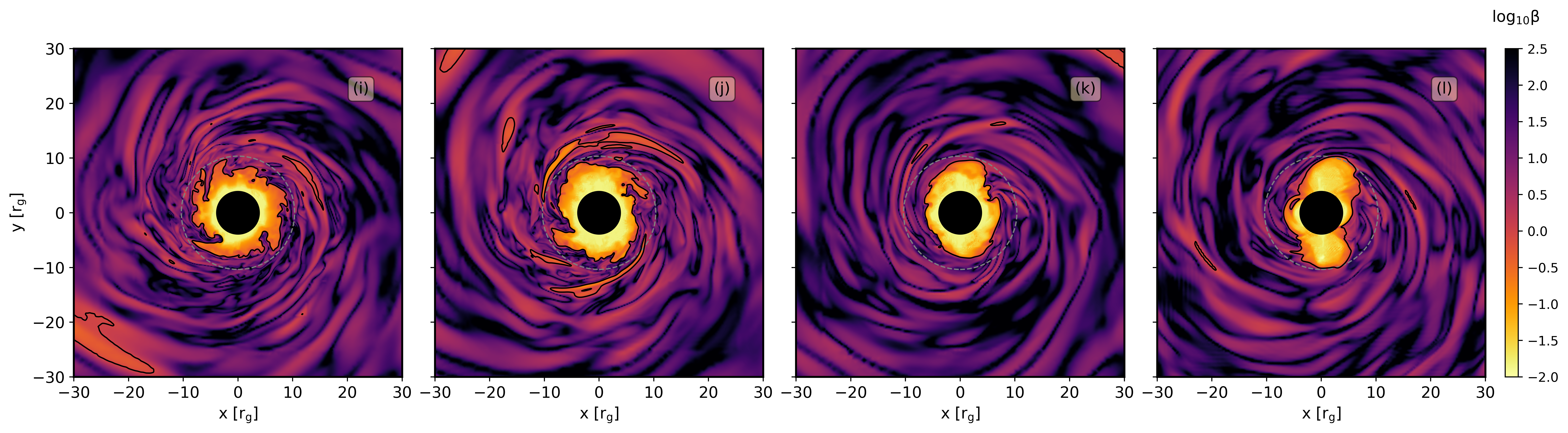}
     \caption{Inner magnetosphere for different rotation and magnetic axis inclinations at t = 12000$r_g/c$. The upper panel shows $log_{10}\rho$ and the middle panel shows $\log_{10}\sigma$ in the z-x plane. The lower panel shows $\log_{10}\beta$ in the equatorial plane (x-y). The gray dashed lines show the corotation radius. The black lines in the lower panel shows the $\beta = 1$ contour. }%
    \label{fig:diffangle}
\end{figure*}

We have carried out five 3D simulations in total. In order to observe the effects of stellar magnetic inclination ($\chi_{\rm star}$) on the accretion dynamics, we performed four different simulations for $\chi_{\rm star} \in [0^{\circ}, 30^{\circ}, 60^{\circ}, 90^{\circ}]$ and $\Omega=0.03$.  
We have also carried out one simulation to explore dynamics in the propeller case for a star with an inclined magnetosphere (discussed in Section \ref{subsec:strippedjets}). 

The onset of the simulations is similar to the 2D case: as we start rotating the star, an Alfv\'en wave is launched from the stellar surface and sweeps across the domain.  Behind the Alfv\'en wave, the solution approaches a force-free equilibrium. Meanwhile, MRI turbulence in the disk causes the inner disk to move inwards until its ram pressure is balanced by the magnetic pressure of the stellar dipole.  Fostered by the anti-parallel configuration of the initial star and disk field configuration, the magnetosphere reconnects with the disk field which opens up initially closed dipolar fieldlines and forms accretion columns. Magnetospheric accretion starts at $\sim 3700 r_g/c$.

Figure \ref{fig:diffangle} shows the inner magnetosphere at a quasi-stationary state for different stellar inclinations with $\mu = 20$ ($\mu$ is the magnetic dipole moment) and $\Omega = 0.03$. The first two rows in Figure \ref{fig:diffangle} show density and magnetization profiles in the z-x plane for different inclinations. The third row shows plasma beta in the equatorial (x-y) plane.  For the aligned case, the star accretes through two almost symmetrical accretion columns in the upper and lower magnetic hemispheres\footnote{Here upper (lower) magnetic hemisphere is the region above (below) the magnetic equator. For $\chi_{\rm star} = 0$, the upper magnetic hemisphere is defined as the region with $z > 0$.}. As we start inclining the stellar magnetic field, similar to the aligned case, the disk opens up the initially closed stellar flux, but now the accretion column further away from the equatorial disk becomes increasingly weaker with higher inclinations and all the matter flows through the column closer to the disk.  

In the inclined cases, it is striking that the dominant accretion column is formed below the upper polar cap for $x > 0$ and above the lower polar cap for $x < 0$. 
In fact, as we will discuss further in Section \ref{subsection:jetformation}, the accretion column moves towards the location of anti-parallel magnetic fields.  This is most evident in the case of $\chi_{\rm star} = 90^{\circ}$ (Figure \ref{fig:diffangle} (h)) which shows the column notably below the equatorial region (for $x>0$).

Tongues of accreting gas form at the disk-magnetosphere boundary due to Rayleigh-Taylor instabilities (RTI) \citep{Arons1976a, Arons1976b, Scharlemann1978, Romanova2003, Kulkarni2008}. These tongues of matter result in accretion via secondary columns close to the equatorial plane (most pronounced in the aligned case).  The instabilities at the disk-magnetospheric boundary allow matter to sweep  between the magnetic fieldlines, resulting in a smaller effective magnetospheric radius compared to the 2D simulations.  For $\chi_{\rm star} \gtrsim 30^{\circ}$ , RTI plays a lesser role as the matter is channeled directly onto the inclined magnetic poles. This was also observed in previous 3D star-disk simulations by \cite{Kulkarni2008}. With increasing stellar magnetic inclinations, the magnetic pole comes closer to the equatorial disk, and it becomes energetically favorable to channel matter directly onto the magnetic poles. As a result of this, we see that the oblique magnetosphere becomes more extended in y-direction (the axis perpendicular to the plane of spin and stellar dipole moment) in the (x-y)-plane.  
\\
\subsection{Jet Formation}\label{subsection:jetformation}
The accretion disk collimates the open stellar magnetic flux to Poynting-flux dominated jets. Here we define the jet as the magnetically dominated region ($\sigma > 1$) around the axis.   
\begin{figure}
   \centering
     \includegraphics[width=8.8cm, angle=0]{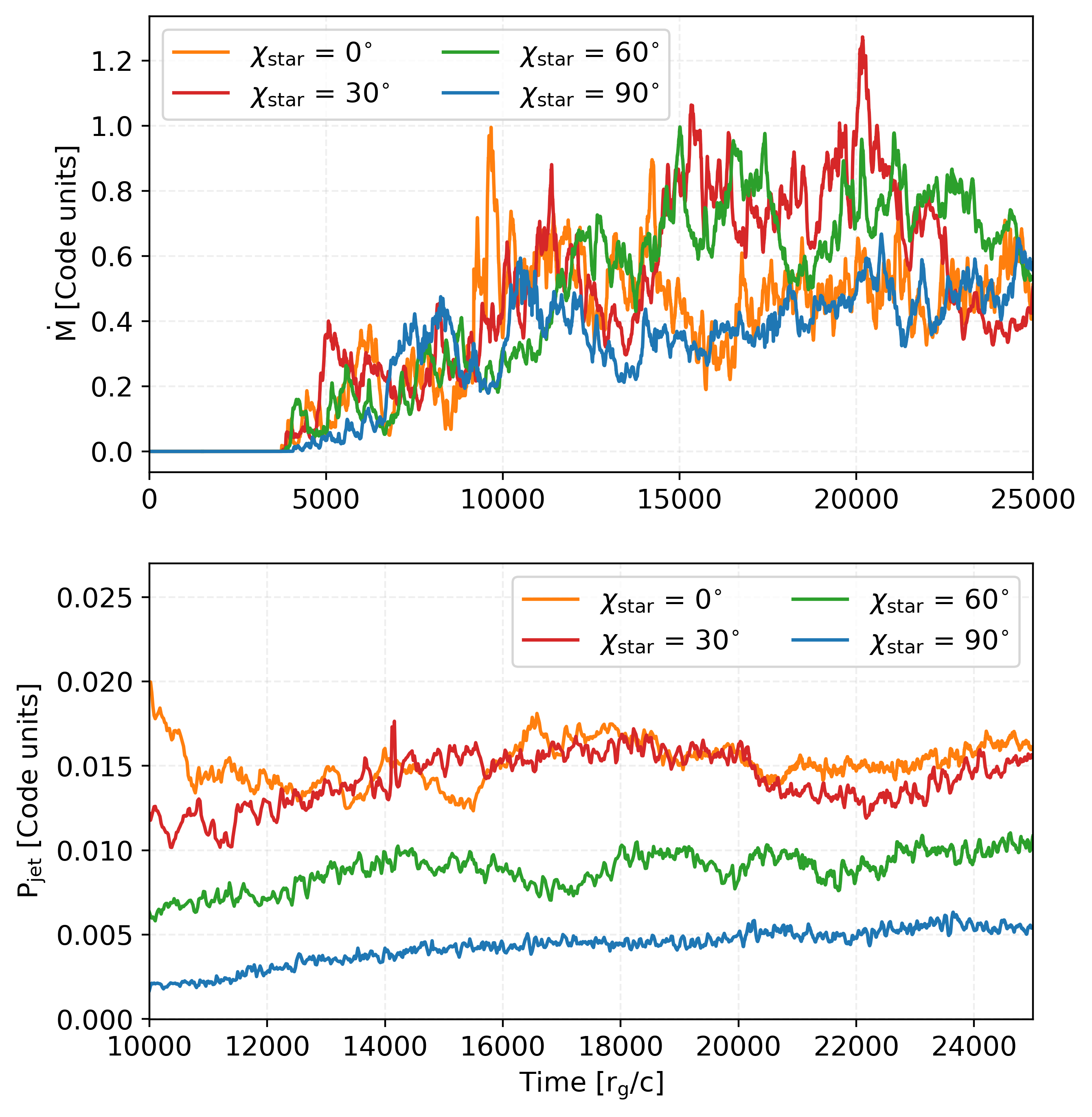}
     \caption{(a) Mass accretion rate at the stellar surface for different stellar magnetic inclinations. (b) Evolution of jet power with time for different magnetic inclinations. The jet is defined as $\sigma = 1$ contour. The mass accretion rate and the jet power is extracted at $4.2 r_g$ and $50 r_g$ respectively.}%
    \label{fig:jet-timeseris}
\end{figure}
Figure \ref{fig:jet-timeseris} shows the time series of the mass accretion rate at the stellar surface and jet power extracted at 50$\rm r_g$ for different stellar inclinations. The mass accretion rate and jet power are defined as follows,
\begin{align}
\centering
\dot{\rm M} &= \int_{0}^{2 \pi} \int_{0}^{\pi} \sqrt{-g} \rho u^r \,d\theta \,d\phi \\
\text{P}_{\rm jet} &= - \int_{0}^{2 \pi} \int_{0}^{\pi} \sqrt{-g} {\text{T}^{\text{EM}}}^r_{t} \,d\theta \,d\phi
\end{align}
where 
\begin{align}
    {\text{T}^{\text{EM}}}^r_{t} = b^2 u^r u_t + \frac{b^2}{2} {\delta^r}_t - b^r b_t
\end{align}
The jet power decreases consistently with increasing magnetic inclination. However, the mass accretion being governed by the disk does not show any such correspondence. For $\chi_{\rm star} \in [0^{\circ},30^{\circ},60^{\circ},90^{\circ}]$, the average mass accretion rate is  $\dot{\rm M} = [0.441, 0.795, 0.708, 0.394]$ with a significant variance in each case ($c_v = \frac{\rm SD (\langle \dot{M} \rangle)}{\langle\dot{M} \rangle} = [0.202, 0.13, 0.173, 0.14]$). The jets formed in all the cases explored here are quite stable and symmetric in both upper and lower hemispheres (for $\sim 25000r_g/c$).

\begin{figure}
   \centering
     \includegraphics[width=8.5cm, angle=0]{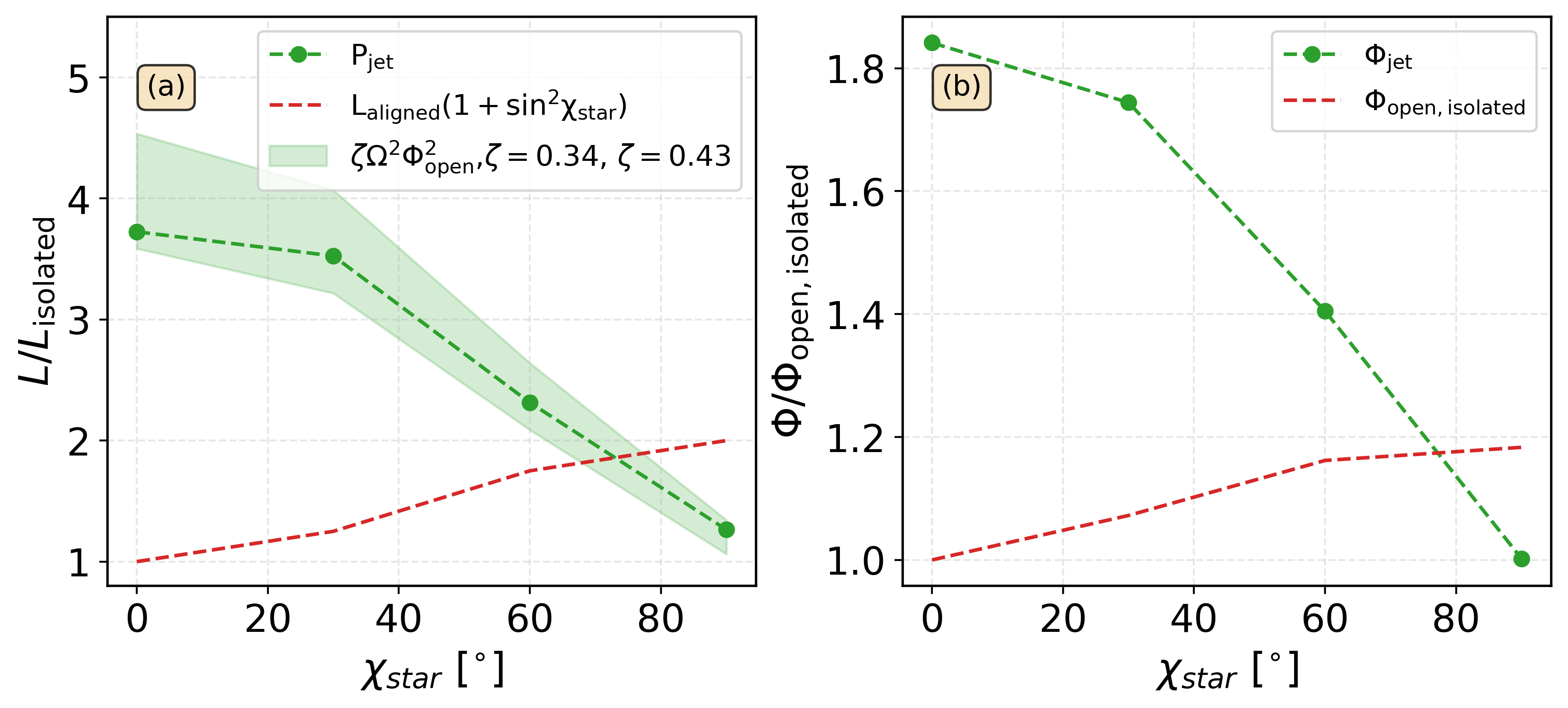}
     \caption{Variation of Jet power (a) and open flux (b) with magnetic axis inclination ($\chi_{\rm star}$). The vertical axis shows the mean quantities normalized by the aligned isolated cases for t $\in$ [15000,25000]$r_g/c$. The red dashed lines represent the variation of power and open flux in the isolated cases. The values of isolated open flux as a function of $\chi_{\rm star}$ are adopted from \cite{Tchekhovskoy2013}.}%
    \label{fig:diffangle-power}
\end{figure}

 Fig \ref{fig:diffangle-power} (a) shows variation of jet power ($\rm P_{jet}$) and panel (b) shows variation in open flux ($\rm \Phi_{open}$) with inclination. All the quantities are averaged over  $t \in$ [15 000, 25 000]$r_g/c$. We also show corresponding quantities for the isolated (non-accreting) case. For accreting neutron stars with aligned rotation and magnetic axes, previous axisymmetric studies suggest that the disk enhances the isolated open flux, leading to enhanced jet power \citep{Parfrey2016, ParfreyTchek2017, Das2022}. The wind power scaling with stellar magnetic inclination for the isolated case is given by $\rm L \simeq L_{aligned} (1 + sin^2\chi_{star}$) \citep{Spitkovsky2006,Tchekhovskoy2013}. However, for neutron star jets, we find that the variation in power with inclination is opposite to that of the isolated wind power scaling. 
 
There are several effects that come into play regarding the jet power of the oblique rotator.  Already the isolated case which was studied in detail by \cite{TchekhovskoyPhilippov2015} is quite subtle.  In essence, by inclining the dipole axis, the distribution of the open magnetic field becomes a non-uniform function of $B^r(\theta,\phi)$ which means, there is no direct correspondence between magnetic flux (surface integral of $B^r$) and Poynting flux (surface integral of $(B^r)^2$).  In \cite{TchekhovskoyPhilippov2015}, around $40\%$ of the power increase was attributed to an overall increased open flux and $60\%$ to the non-uniform ``bunching'' of the open magnetic field lines.  

In our simulations, inclination-dependent equatorial field line bunching plays only a minor role. This is because non-uniform bunching of isolated MHD winds is sub-dominant against the collimating effect from the pressure of the disk. A quantitative comparison of the distribution of Poynting flux and radial magnetic field across the jet is provided in Appendix \ref{appendix2}.

Another mechanism that affects the jet power is the disk-induced magnetic flux opening. Similar to previous 2D simulations \cite{Das2022}, we initialize our domain with an anti-parallel star-disk magnetic field configuration. In the aligned case, the amount of open magnetic flux is then given by the location of the magnetospheric radius in the equatorial plane \citep{Parfrey2016,ParfreyTchek2017,Das2022}. However, in the 3D case, disk-induced flux opening becomes less efficient with magnetic obliqueness. This is consistent with the increasingly larger closed zone seen also in the middle row of Figure  \ref{fig:diffangle}. 
Overall, for the accreting cases presented here, the open magnetic flux varies by a factor $\approx 2$ from the aligned case to the orthogonal rotator.  

To a good approximation, the jet power in our simulations is related to the open magnetic flux as $L=\xi\Omega^2\Phi_{\text{open}}^2$ with $\xi \in [0.34,0.43]$ which is indicated by the green shaded area in Figure \ref{fig:diffangle-power}.  This suggests that disk-induced flux opening determines the jet power which leads to a decrease in power with increasing obliqueness.  

We are hence left with one question: what is the reason for the reduced flux opening efficiency for the oblique case?  
The answer must be sought in the 3D geometry of the star-disk system. In the aligned configuration, disk-connected field lines are subjected to ballooning and flux opening on both hemispheres \citep[e.g.][]{Uzdensky2004}.  The newly opened field lines are subsequently collimated into a (northern- and southern-) jet by the lateral pressure of the accretion disk.  
In the highly oblique case, the situation changes qualitatively: since the poles point towards the direction of the disk, a larger fraction of already open magnetic field lines are able to reconnect with the disk fields. The sketch in Figure \ref{fig:cartoon} shows the contrasting scenarios for the aligned case and the orthogonal rotator. In the former case, two newly opened field lines are created, one contributing to the northern- and one to the southern- jet.  In the latter case, no additional open flux is obtained when the disk reconnects with already open field lines (green boxes in Figure \ref{fig:cartoon}), however the reconnected field line can move across the disk to end up in the jet on the other hemisphere.  
Since the accretion columns are bounded by the last open field line -- one of which has just moved across the disk -- we can now understand why the accretion columns shown in Figure \ref{fig:diffangle} are tilted towards the anti-parallel region of the star-disk system (top-left and bottom right quadrants).  
Also in the case of the orthogonal rotator, the closed dipolar stellar field lines are able to reconnect with the disk fields and open up. This situation is highlighted by the two inner purple boxes in panel (d) of Figure \ref{fig:cartoon}.  In our simulation however, this flux opening is almost exactly compensated by the absorption of initially open flux by the equatorial accretion disk.  Thus in the case of the orthogonal rotator, we end up with zero net flux opening (cf. Figure \ref{fig:diffangle-power}).  

\begin{figure}
    \centering
    \text{$\chi_{\rm star} = 30^{\circ}$} \\
    \begin{tabular}{c c}
        \includegraphics[width=4.2cm, angle=0]{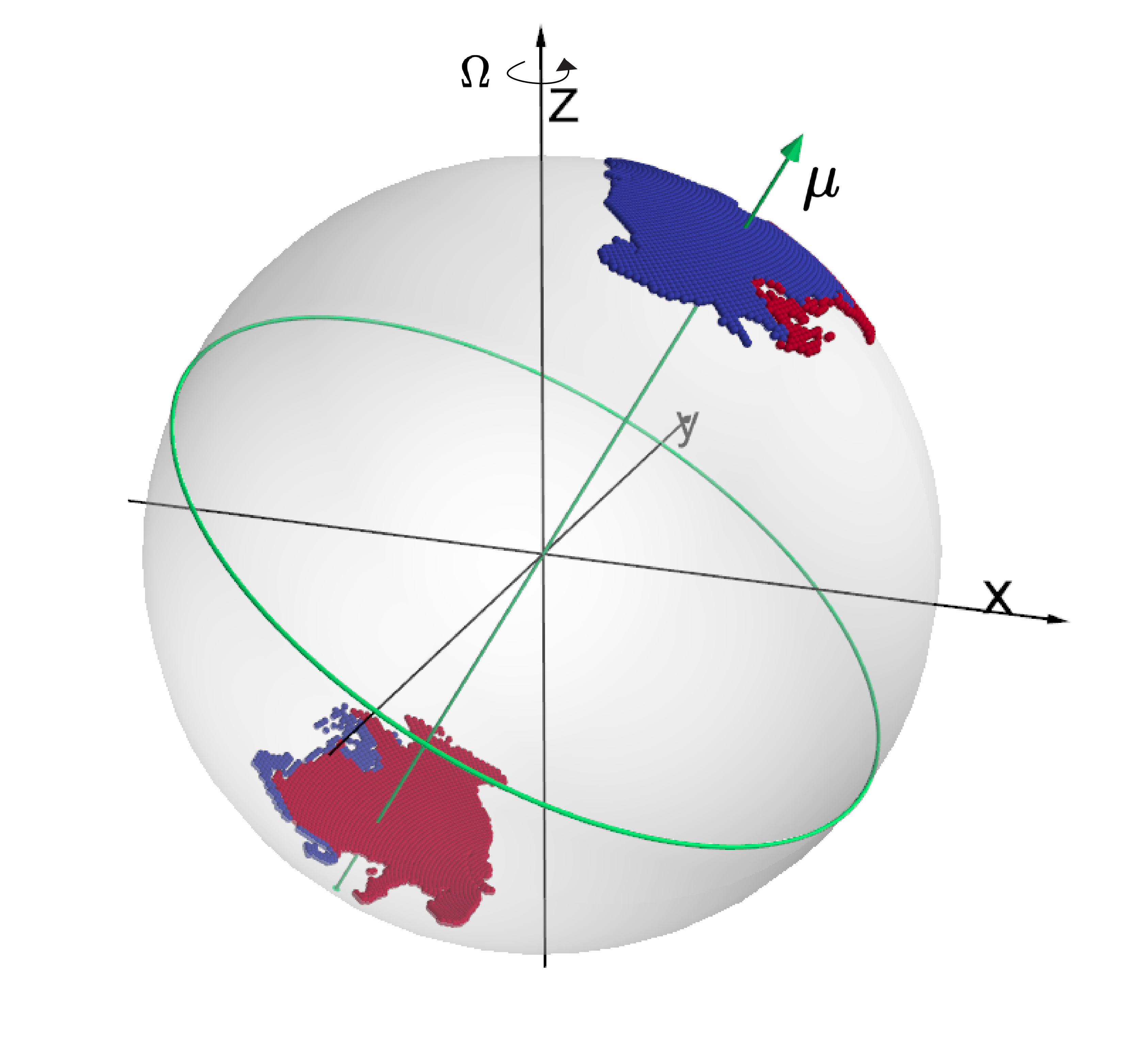}\hspace{-0.5cm} &
        \includegraphics[width=4.2cm, angle=0]{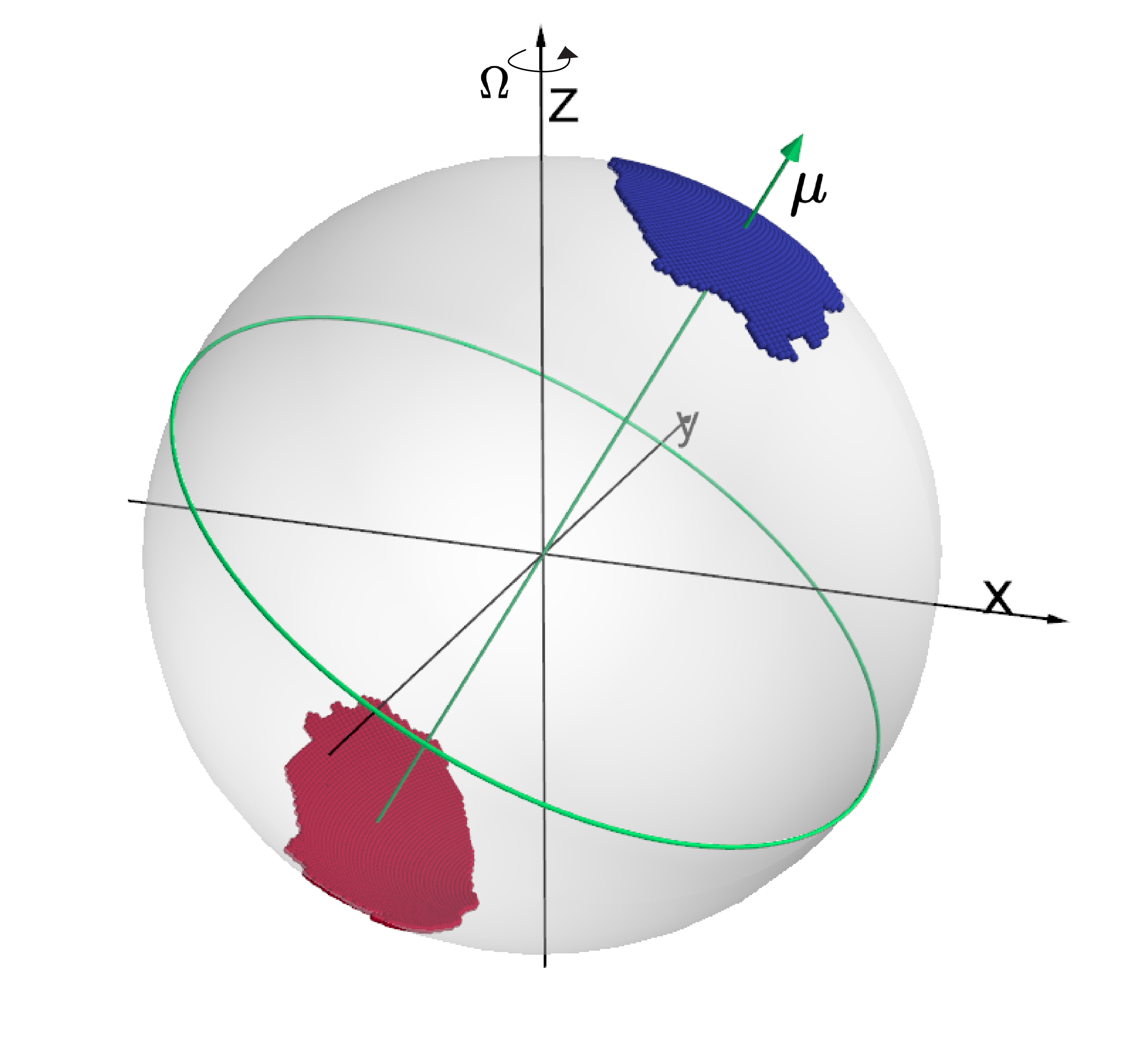}\hspace{-0.5cm} \\
    \end{tabular}
    \text{$\chi_{\rm star} = 60^{\circ}$} \\
    \begin{tabular}{c c}
        \includegraphics[width=4.2cm, angle=0]{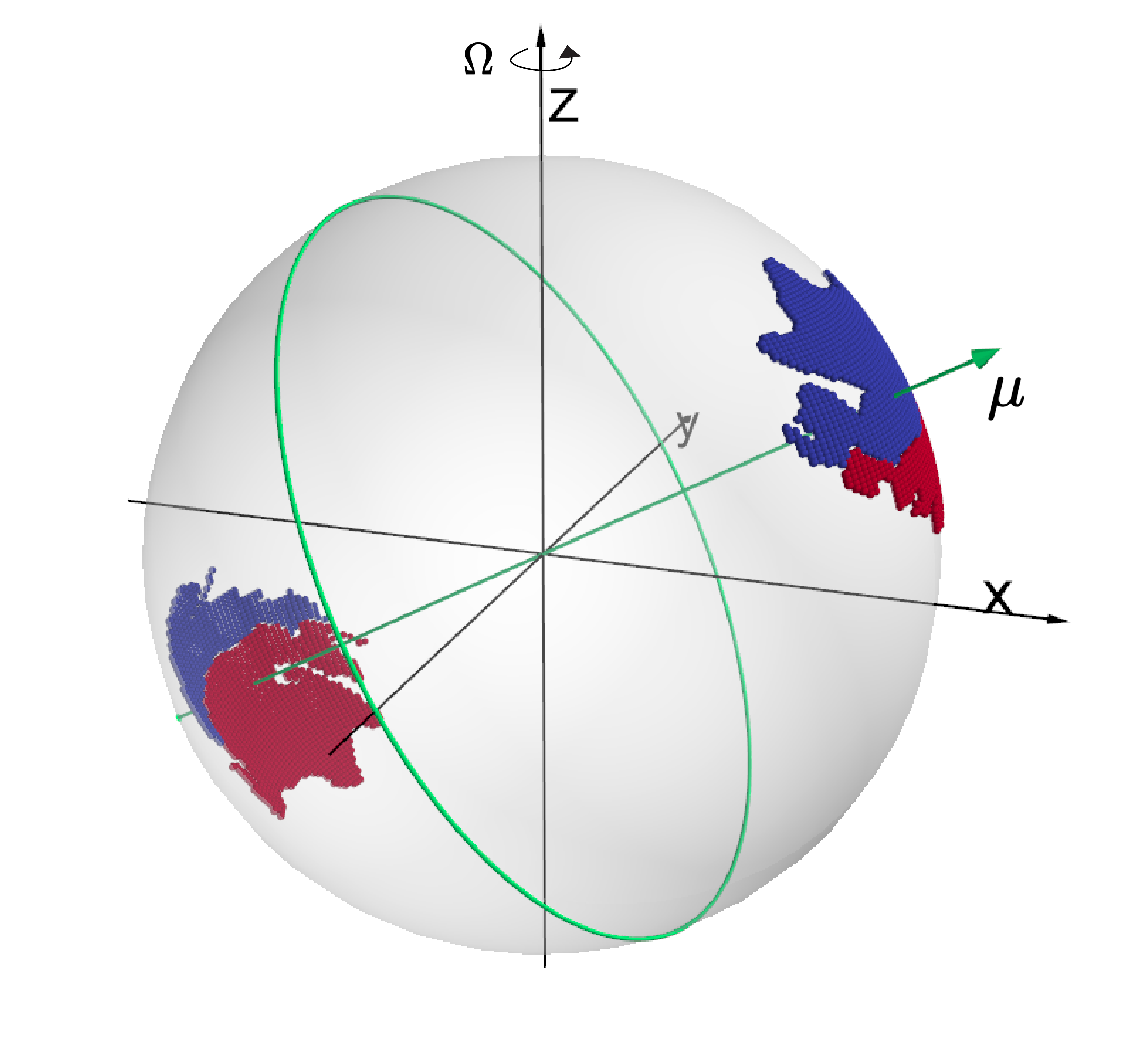}\hspace{-0.5cm} &
        \includegraphics[width=4.2cm, angle=0]{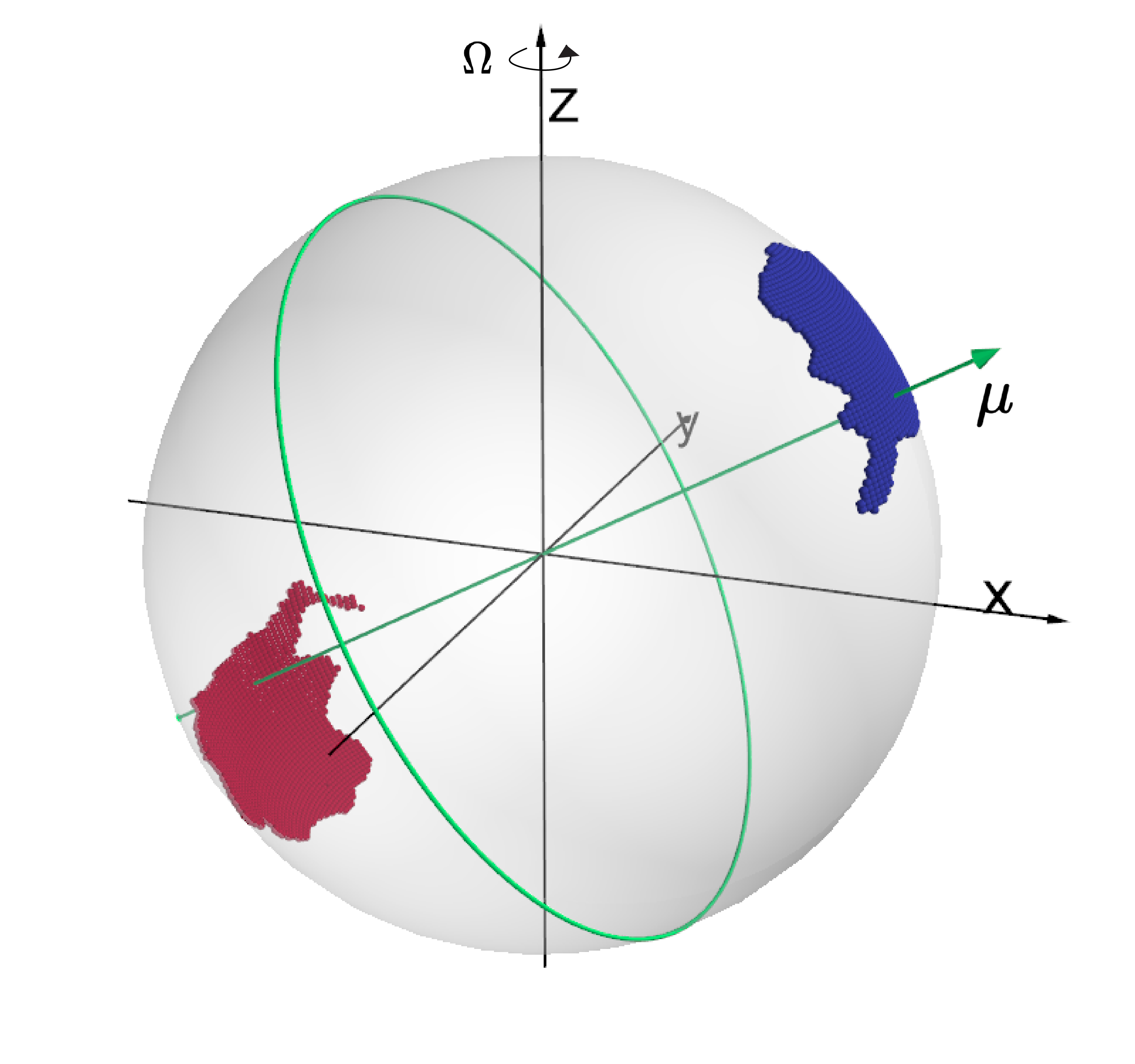}\hspace{-0.5cm} \\
    \end{tabular}
    \text{$\chi_{\rm star} = 90^{\circ}$} \\
    \begin{tabular}{c c}
        \includegraphics[width=4.2cm, angle=0]{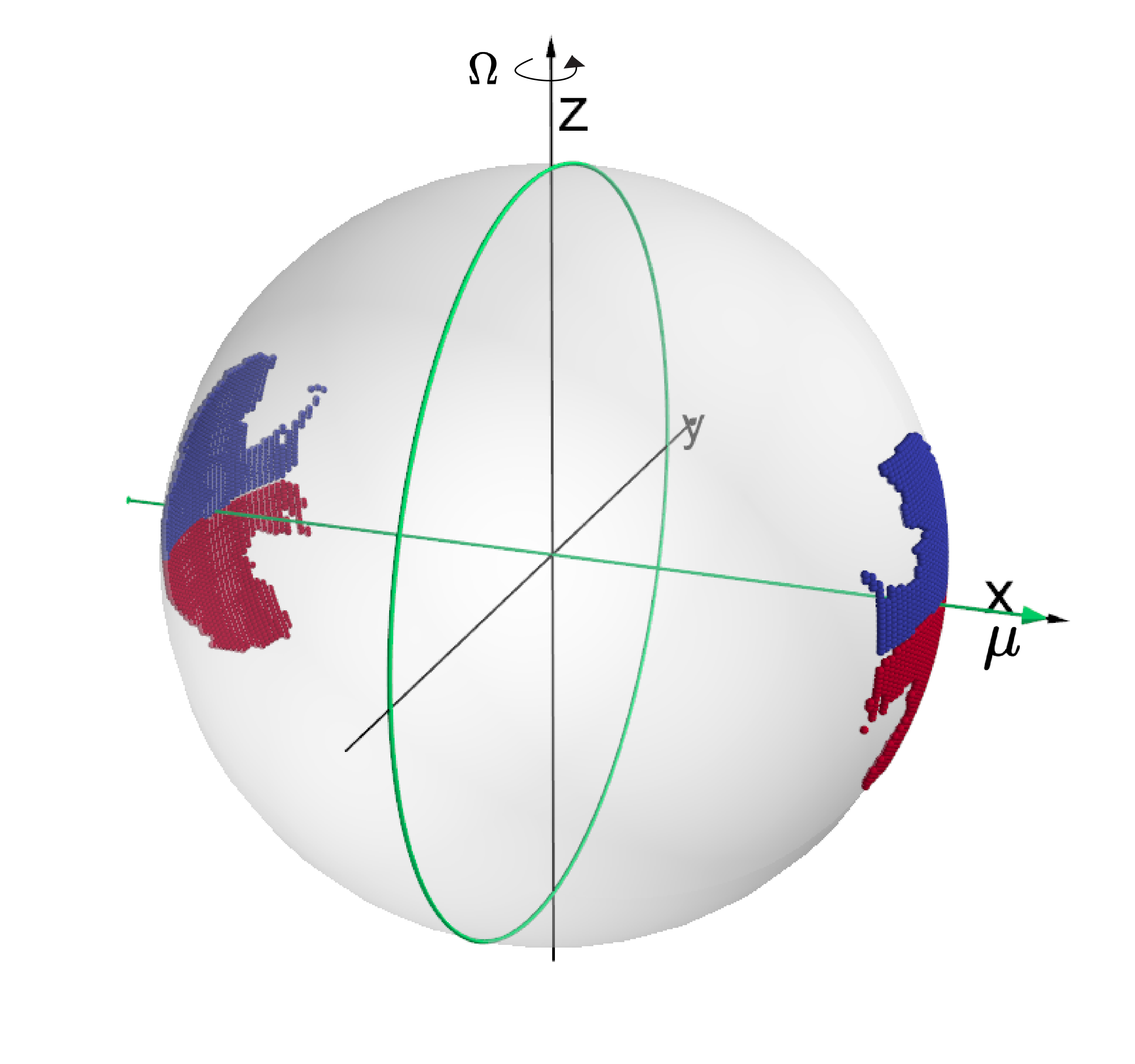}\hspace{-0.5cm} &
        \includegraphics[width=4.2cm, angle=0]{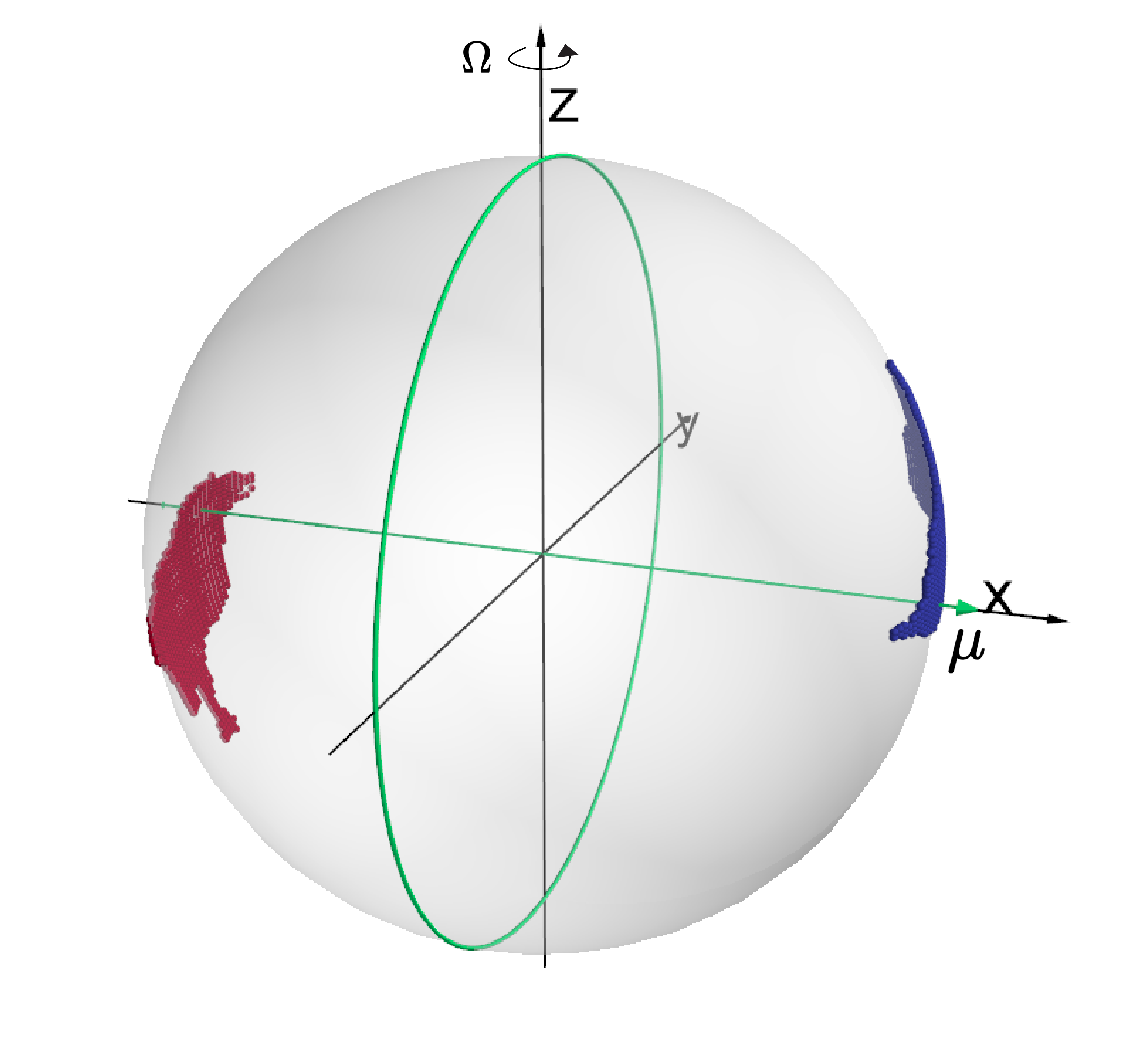}\hspace{-0.5cm} 
    \end{tabular}
    \begin{tabular}{c c}
    \text{\hspace{2.4cm}(a)\hspace{1.7cm}} & \text{\hspace{1.9cm}(b)\hspace{2cm}} \\
    \end{tabular}
    \caption{Surface maps of the open magnetic fieldlines powering the jet for different stellar magnetic axis inclinations in the accreting regime. The blue and red colors represent the fieldlines ending up in the jet in the northern ($z > 0$) and southern hemispheres ($z < 0$) respectively. The first column (a) represents the magnetic foot-points of the open fieldlines at an initial time (mimicking the isolated case) and the second column (b) shows the foot-points of the open field lines in the jet at a later time (after reaching the quasi-stationary state). The green arrow indicates the direction of the magnetic moment. For a dipolar geometry, the polarity of the magnetic field flips at the magnetic equator (outlined by the green circle), $b^r > 0$ (out-going fieldlines) in the hemisphere along the magnetic moment and $b^r < 0$ (in-going fieldlines) in the rest of the hemisphere.}
    \label{fig:trac-surfacemap}
\end{figure}

Efficient reconnection of the anti-parallel fields leads to the situation that the simulated jets are fed by only one pole per hemisphere even for stars with extreme obliqueness of $90^\circ$.  
To further elucidate this point, we show the footpoints of the open magnetic field in Figure \ref{fig:trac-surfacemap}.  
In the left column (a), we show the initial wind-like configuration recovered in our simulations before the onset of MRI-driven accretion.   In the right column (b) we illustrate the situation in the late stage of the simulation.  The colors represent foot-points of the fieldlines contributing to the upper (blue) and lower (red) jet.  In practice, field-lines are traced up to $r > 50r_g$ and are tagged as part of the jet as long as they remain in magnetically dominated $\sigma>1$ regions.  

For an isolated pulsar, in the aligned case ($\chi_{\rm star} = 0^{\circ}$), for a given rotational hemisphere, the open-flux consists of fieldlines that connect to only one magnetic pole. But with increasing magnetic inclination, more fieldlines of the wind connect to the ``other'' pole which gives rise to the striped wind \citep{Michel1971}.  This case is illustrated in Figure \ref{fig:trac-surfacemap}a). On the surface of the star, we tag footpoints of field lines that end up in the upper (rotational) hemisphere ($z>0$) as blue and as red if they end up in the lower hemisphere ($z<0$).  With increasing inclination, more field lines with opposite polarity are found in either hemisphere (hence more admixture of red and blue tagged footpoints respectively).  After the onset of accretion,  the disk absorbs some of the initial open flux and even moves open field lines across the disk (as sketched in Figure \ref{fig:cartoon}).  Ultimately this leads to a reduction of flux opening efficiency and a single polarity outflow in either hemisphere (Fig \ref{fig:trac-surfacemap}b).

During the review process, we came across another set of 3D GRMHD accreting neutron star simulations which explores the stellar magnetic inclination parameter space by \citep{Murguia-Berthier2023}. They also report the same trend in jet-power with magnetic inclination for the anti-parallel star-disk geometry for their $\mu = 10-20$ cases.

\subsection{Propeller Jets}\label{subsec:strippedjets}

Along with a parameter sweep on the magnetic inclination angle, we have also performed one run to explore the dynamics in a strong propeller regime ($r_m \simeq 0.7r_{lc},\ \simeq 1.9r_{co}$). We initialize our simulation with a similar setup as discussed in section \ref{sec:setup} ($\mu = 30, \Omega = 0.05$) and set the stellar magnetic inclination to $60^{\circ}$. In order to maintain numerical stability in this more challenging scenario, the region inside the light- cylinder is initialized with $\sigma_t = 60$ and $\beta_t = 0.016$ and the pressure and densities are floored such that  $\beta > 0.003$ and $\sigma < 300$. 

\begin{figure}
    \centering   
     \includegraphics[width=8.8cm, angle=0]{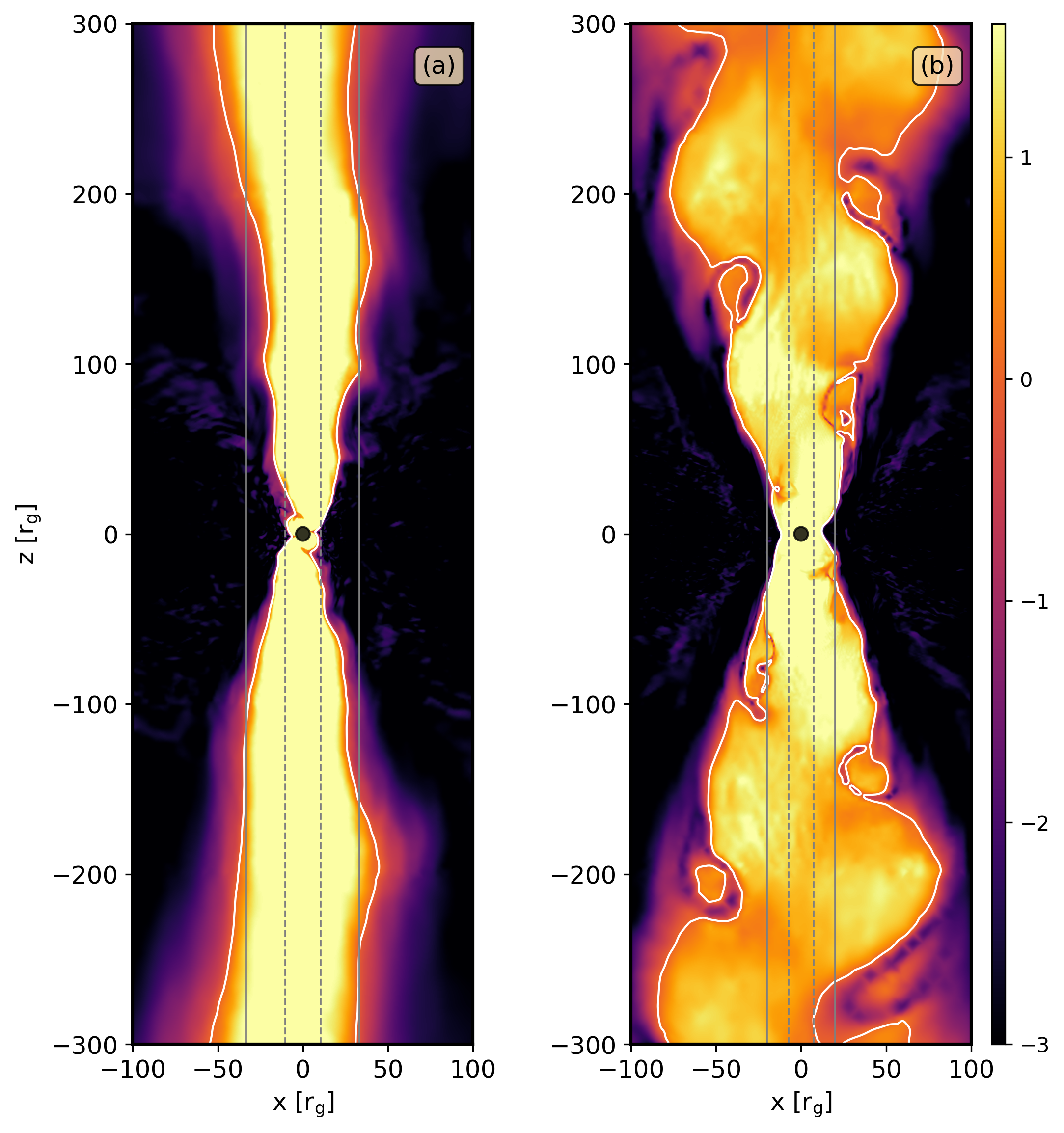}
     \caption{Jet structure in the z-x plane for (a) accreting case with $\chi_{\rm star} = 60^{\circ}, \mu = 20$ and $\Omega = 0.03$ (b) strong propeller with $\chi_{\rm star} = 60^{\circ}, \mu = 30$ and $\Omega = 0.05$. The color-bar shows log$_{10} \sigma$ and the white lines show the $\sigma = 1$ contour. The gray dashed and solid lines show corotation radius and the light cylinder radius respectively. }
     \label{fig:stripped-jet}
\end{figure}

The morphology of the jets for $\chi_{\rm star} = 60^{\circ}$ is shown in Figure \ref{fig:stripped-jet} for the accretion regime (a) and propeller regime (b). Similar to the previous cases, the disk collimates the open fieldlines to form a jet. However, there are some differences in the jet structure in these two regimes. First, due to the larger magnetospheric radius, the jet opening angle in the propeller regime is larger compared to the accreting cases. 
Second, similar to the striped wind from isolated oblique rotators, the misaligned propeller shows current sheets within the jet.  The current sheets first occur once the flow crosses the light cylinder but, since the entire flow is collimated by the ambient disk pressure, they are not restricted to the traditional striped wind region located $\pm\chi_{\rm star}$ around the equator.  
\begin{figure}
\centering
    \text{$\chi_{\rm star} = 60^{\circ}$} \\
    \begin{tabular}{c c}
        \includegraphics[width=4.2cm, angle=0]{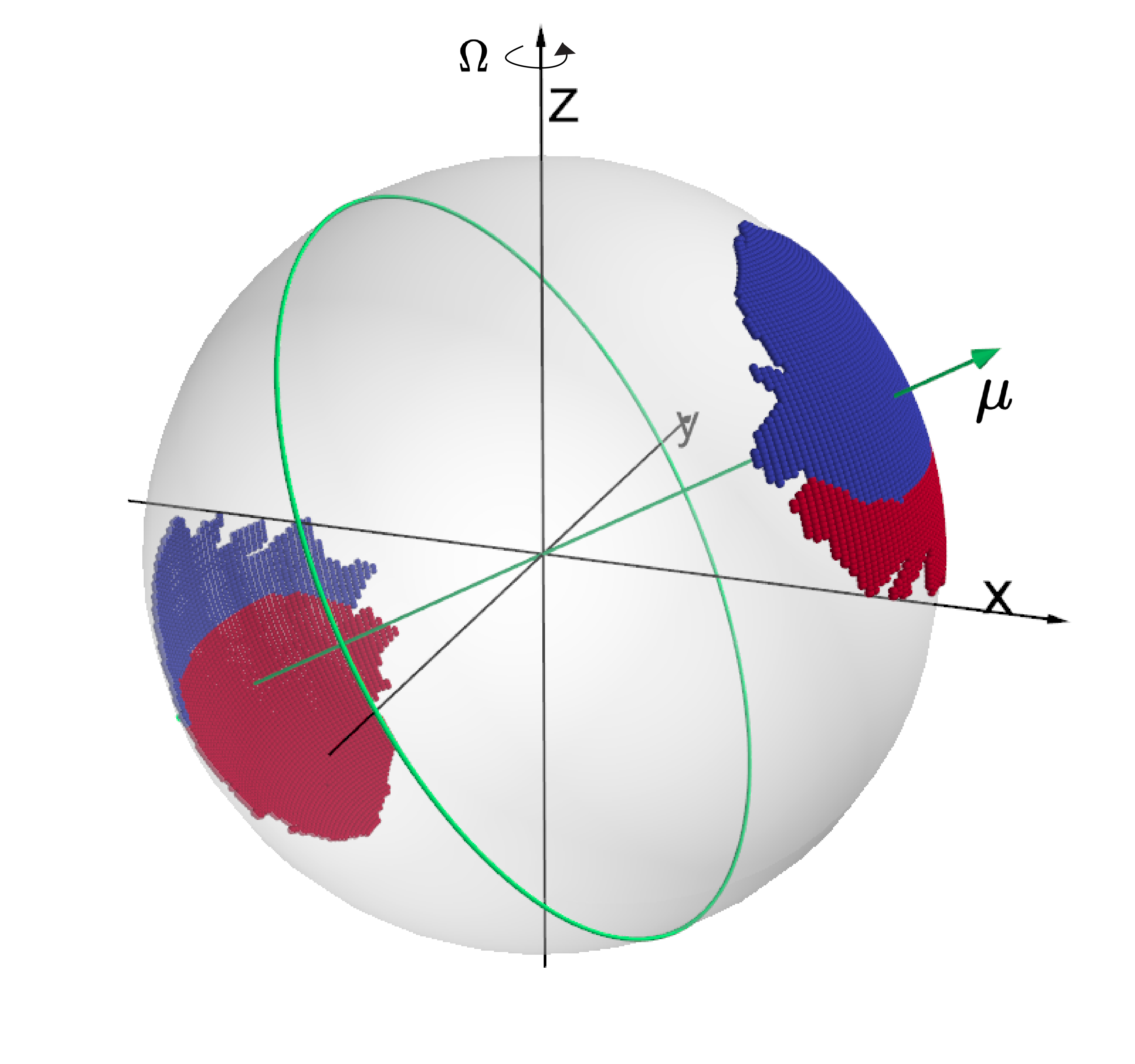}\hspace{-0.5cm} &
        \includegraphics[width=4.2cm, angle=0]{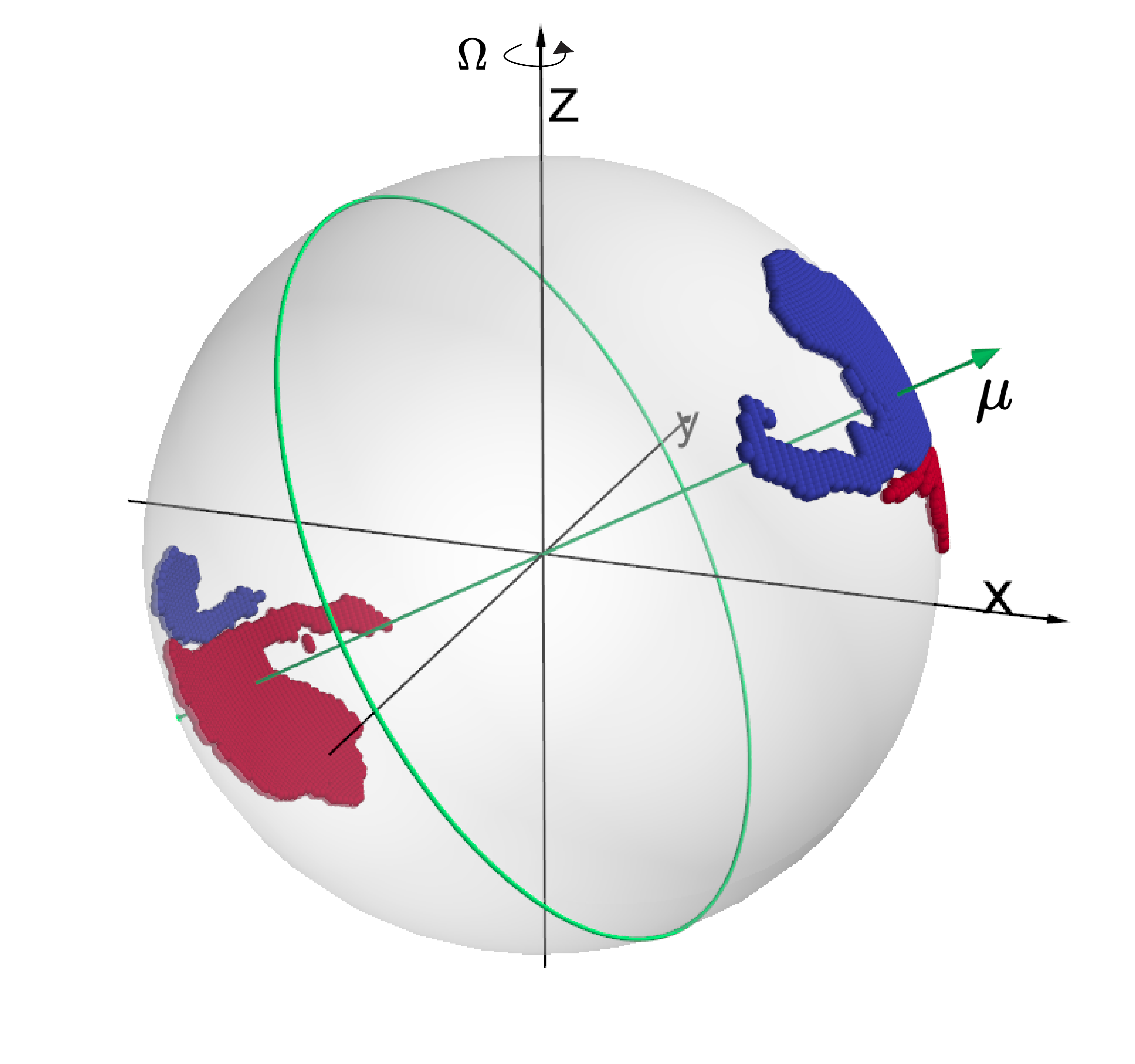}\hspace{-0.5cm} \\
    \end{tabular}
    \begin{tabular}{c c}
    \text{\hspace{2.4cm}(a)\hspace{1.7cm}} & \text{\hspace{1.9cm}(b)\hspace{2cm}} \\
    \end{tabular}
\caption{Surface maps of the open magnetic fieldlines powering the jet in the upper (blue) and lower (red) hemispheres. Here $\chi_{\rm star} = 60^{\circ}$, $\mu = 30$ and $\Omega = 0.05$. The two panels show the surface maps at two different times (a) Early (b) Quasi-stationary state. All the labels are the same as in Fig \ref{fig:trac-surfacemap}.}
\label{fig:trac-surfacemap-prop}
\end{figure}
In our highly oblique accreting scenarios, stripes are also observed at early times, but disappear once the fieldlines in anti-parallel regions reconnect with the disk.  At large distances, the jet structure is similar to the aligned case. 
In the strong propeller regime, reconnection between the open polar fieldlines of the star and the disk magnetic fields (green boxes in Figure \ref{fig:cartoon}) is inhibited. As a result, the stripes survive within the jet. This is further illustrated by a field line foot-point map (Figure \ref{fig:trac-surfacemap-prop}). Unlike the accreting cases, here we can see that at a quasi-stationary state, the jet has two polarity fieldlines in both hemispheres and hence the stripes in the jet. 

\section{Discussion \& Conclusion}

Our simulations should be applicable to the hard state of accreting millisecond pulsars.  They show that the formation of jets in this parameter regime is quite robust, even for highly oblique cases of the star-disk magnetospheric system.  
Compared to black hole jets where the entire magnetic field must be sourced outside of the event horizon, the inclination of the \textit{internal} field of the neutron star offers an additional free parameter.    Hence for otherwise identical initial conditions and for the same mass accretion rate, we observe a reduction of jet power by a factor of $\simeq 2.95$ between the orthogonal rotator and the aligned configuration.  This is explained by the less efficient flux opening of the inclined star-disk system.  
Assuming that neutron star jets are indeed formed by the mechanism studied here, unless all neutron star systems conspire to the same obliqueness angle, we thus expect their jets to be more widely scattered in power than black hole jets.  Recent high-sensitivity radio observations lend some support to this prediction \citep{vandenEijnden2021}.  At the same time, the jet power in all models presented here is sufficiently strong to explain radio emission from both weak- and strong magnetic field sources such as Cir X-1, Sco X-1 and Swift J0243.6+6124 \citep[e.g.][]{Fender2004, BradshawFomalontEtAl1999}.  
In particular, scaled to physical units and assuming a polar magnetic field strength of $10^9 \rm G$, the $\chi_{\rm star}=90^\circ$ case (the weakest jet presented here) corresponds to a jet power of $\simeq8\times 10^{36}\rm erg\, s^{-1}$, almost two orders of magnitude above the minimum jet power estimated by \cite{Fender2004} for Cir X-1.  
We refer the reader to \cite{Das2022} for further discussion of the jet power scaling.  

To compare with the earlier 2D results, we have performed one 2D simulation with matched initial conditions to the $\chi_{\rm star}=0^\circ$ case shown here \footnote{The numerical setup for the 2D run is otherwise the same as in \cite{Das2022}}. In 3D simulations, due to Rayleigh-Taylor instabilities developing at the disk-magnetospheric boundary, the magnetospheric radius for $\chi_{\rm star} = 0^{\circ}$ is much closer to the star compared to the 2D case. However, the mass accretion rate in both cases are quite similar, 0.318 and 0.279 (code units) for 2D and 3D respectively. The variability in the mass accretion rate in the 2D ($c_v = 0.38$) simulations is quite large compared to the 3D ($c_v = 0.154$) simulations. This occurs because in 3D geometry, the accretion flow can always penetrate the magnetosphere in between the magnetic fieldlines leading to a comparably stable surface accretion rate. Similar to the mass accretion rate, the jet power in both these cases is also quite close. The average jet power in 2D and 3D are 0.011 and 0.015 respectively (both in code units). To summarize, for aligned accreting neutron stars, although the accretion dynamics in the inner magnetosphere in 3D change compared to the axisymmetric simulations, these variations are not transferred into the jet. 

Another possible factor that might impact the behavior of jets in the 3D simulations is the dependence of flux opening on the initial star-disk magnetic field geometry. In all of our simulations, we have considered anti-parallel star-disk magnetic fields which in principle allows the disk magnetic field to reconnect with the stellar magnetic field which leads to enhanced flux opening (for lower inclinations). Recent 3D GRMHD simulations of the aligned case by \cite{Parfrey2023} suggest that for $r_m \ll r_{co}$, the difference between parallel and anti-parallel star-disk geometry is quite prominent, with the parallel geometry resulting in less jet-power compared to the anti-parallel case. However, the disk-polarity dependence is reduced for higher inclinations and when $r_m \simeq r_{co}$ such that simulations with both disk-polarities result in comparable jet powers \citep{Parfrey2023, Murguia-Berthier2023}.



The simulation of the \textit{propeller} regime shows that the neutron star jets resulting from an inclined star can contain stripes of magnetic field with opposite polarity. 
In general, the stripes are possible if the jet is comprised of open field lines that originate from opposite magnetic poles.   In the isolated case, the striped wind geometry is well understood and confined to the angular range $\pi/2 - \chi_{\rm star} < \theta < \pi/2 + \chi_{\rm star}$. Since the disk collimates the jet however, we find that the stripes in the accreting scenario can preside much closer to the polar axis (for $\chi_{\rm star} = 60^{\circ}$ within $8^{\circ} < \theta < 20^{\circ}$ and $160^{\circ} < \theta < 172^{\circ}$). 

Even though the striped jets only arose in the strong propeller simulation, striped jets might survive also in accreting regimes. As long as opposite polarity fieldlines are collimated into a jet of one hemisphere, current sheets are expected to form within the outflow. This might also happen for the accreting case depending on the reconnection efficiency of the star-disk magnetic field, the stellar magnetic inclination, and the initial magnetic field in the disk. 

The possibility of striped jets is interesting since it provide a means for magnetic energy dissipation via magnetic reconnection in the jet. For isolated stars with misaligned spin and magnetic axis, the dissipation in the striped wind has been studied by many authors \citep{Coroniti1990, Lyubarsky2003, Komissarov2013}, most recently by  \cite{Cerutti2020, Hakobyan2023} who find that the global structure of the stripped wind follows the split-monopole prediction \citep{Bogovalov1999} and in the stripped region, plasmoid dominated magnetic reconnection leads to efficient magnetic energy dissipation throughout the wind. In the case of misaligned stars, assuming a split-monopole magnetic wind, the dissipation for $r > r_{\rm lc}$ is given by $L(r) = L_0(1 - \beta_{\rm rec} {\rm ln}(r/r_{\rm lc}))$. Here, $\beta_{\rm rec}$ is the reconnection rate, and $L_0$ is the Poynting flux in a dissipation-less split-monopole magnetosphere ($L_0 = 2cB_{\rm star}^2 r_{\rm star}^4/3r_{\rm lc}^2$) \cite{Cerutti2020}. In our case, for $\chi_{\rm star} = 60^{\circ}$ due to the flux collimation by the accretion disk, the stripes only survive within $8^{\circ} < \theta < 20^{\circ}$ and $160^{\circ} < \theta < 172^{\circ}$. For a given reconnection speed  $\beta_{\rm rec}$, we can estimate the dissipation of the current sheets in the jets following \cite{Cerutti2020}.  With the above quoted extent of the striped region, this leads to $L(r) = L_0(1 - 0.0033 \beta_{\rm rec} {\rm ln}(r/r_{\rm lc}))$.  
For $\beta_{\rm rec} \sim 0.1 - 0.2$, the resulting dissipation radius is very large ($> 10^{30} r_g$) which indicates that dissipation of stripes is dynamically unimportant.  Never the less, the formation of current-sheets in the magnetically dominated jet offers a mechanism for particle acceleration which might power flaring emission in the ``striped state''.   The striped state should emerge when star-disk coupling is moderate: e.g. in the propeller regime when the accretion rate is low and the disk is geometrically thick and able to collimate a jet.  In addition, the reconnection of the star-disk system is reduced for parallel field configurations which could give rise to striped jets even for strongly accreting systems.  

The 3D simulation presented here mark a first step to understand neutron star jet formation from oblique systems.  More work is needed to address the dependence of the jet properties with the parameters of the system, foremost the magnetospheric radius and initial magnetic field geometry. 

\begin{acknowledgments}
PD and OP acknowledge funding from the Virtual Institute for Accretion (VIA) within NOVA (Nederlandse Onderzoeksschool voor Astronomie) Network 3 `Astrophysics in extreme conditions'. Simulations have been carried out in part on the Dutch national e-infrastructure with the support of SURF Cooperative (project number NWO-2022.019), Swedish cluster Dardel as a part of PRACE (DECI-17) allocation (project number 17DECI0051) and the HELIOS cluster of the Anton Pannekoek Institute for Astronomy. We thank Anna Watts and Nathalie Degenaar for valuable comments. 
\end{acknowledgments}

%

\vspace{5mm}
\facilities{\textit{Snellius, Dardel}}


\software{\bhac \citep{Porth2017},
          Python \citep{Python2007},
          NumPy \citep{Numpy2011},
          Scipy \citep{Jones2001},
          MPI \citep{mpi4py},
          Matplotlib \citep{Hunter2007, matplotlibv2},
          Mayavi \citep{Mayavi},
          IPython \citep{IPython2007}}

\bibliography{References,References-new,more-ref}{}
\bibliographystyle{aasjournal}


\appendix
\section{Corotating Cartesian Schwarzschild Coordinates}\label{appendix1}
The line element of the Schwarzschild coordinates in standard spherical coordinates ($t,r,\theta,\phi$) can be written as,
\begin{multline}
    ds^2 = - \bigg(1 - \frac{2M}{r}\bigg)\ dt^2 + \frac{1}{\bigg(1 - \frac{2M}{r}\bigg)}\ dr^2 + r^2\ d\theta^2\\ + r^2\ {\rm \sin^2}\theta\ d\phi^2
\end{multline}
With the following transformations,
\begin{align*}
    x &= r\ {\sin\theta}\ {\cos}(\phi - \Omega t)\\
    y &= r\ {\sin\theta}\ {\sin}(\phi - \Omega t)\\
    z &= r\ {\cos \theta}\\
\end{align*}
the covariant components of the cartesian corotating Schwarzschild metric can be written as,
\begin{align*}
    g_{tt} &= \bigg(1 - \frac{2M}{r}\bigg) + (x^2 + y^2)\ \Omega^2\\
    g_{tx} &= g_{xt} = -y\ \Omega\\
    g_{ty} &= g_{yt} = x\ \Omega\\
    g_{tz} &= g_{zt} = 0\\
    g_{xx} &= \frac{1}{ 1 - \frac{2M}{r}} \bigg(1 - \frac{2M}{r^3}(z^2 + y^2)\bigg)\\
    g_{yy} &= \frac{1}{ 1 - \frac{2M}{r}} \bigg(1 - \frac{2M}{r^3}(z^2 + x^2)\bigg)\\
    g_{zz} &= \frac{1}{ 1 - \frac{2M}{r}} \bigg(1 - \frac{2M}{r^3}(x^2 + y^2)\bigg)\\
    g_{xy} &= g_{yx} = \frac{xy}{1 - \frac{2M}{r}} \bigg(\frac{2M}{r^3}\bigg)\\
    g_{xz} &= g_{zx} = \frac{zx}{1 - \frac{2M}{r}} \bigg(\frac{2M}{r^3}\bigg)\\
    g_{yz} &= g_{zy} = \frac{yz}{1 - \frac{2M}{r}} \bigg(\frac{2M}{r^3}\bigg)
\end{align*}
Here $\Omega$ is the angular corotation frequency. In our simulations, this is always set to $\Omega_{\rm star}$.  

\section{Azimuthally averaged Theta profiles}\label{appendix2}
\begin{figure*}[ht]
    \centering   
     \includegraphics[width=18cm, angle=0]{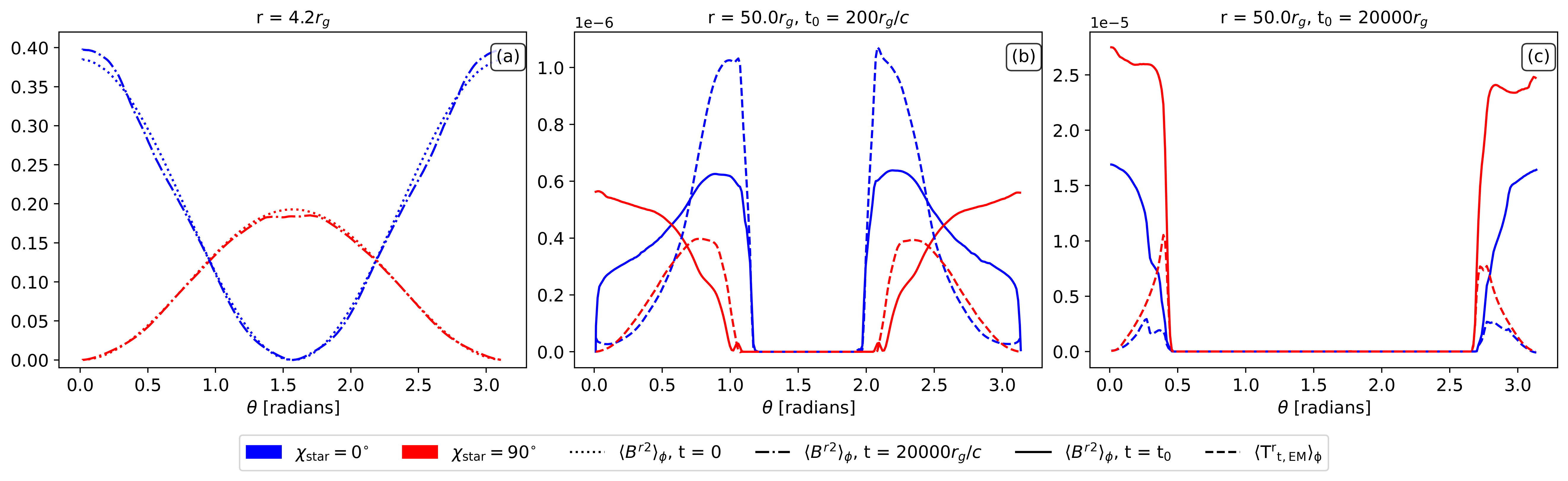}
     \caption{$\theta$ profiles of different quantities for $\chi_{\rm star} = 0^{\circ}$ (blue) and $90^{\circ}$ (red). Panel (a) shows the time evolution of the quantities at the stellar surface. Panel (b) and (c) show the time evolution at $50r_g$ in the regions with $\sigma > 1$. The dotted, dot-dashed lines in (a) show $\langle{B^r}^2\rangle_{\phi}$ at t $\in [0, 20000] r_g/c$. The solid and dashed lines in panel (b) and (c) show $\langle{B^r}^2\rangle_{\phi}$ at t = t$_0$ and $\rm \langle{{T^r}_t}_{,EM}\rangle_{\phi}$ respectively. Here t$_0 = 200 r_g/c$ (b) and t$_0 = 20000 r_g/c$ (c).}%
     \label{fig:theta-profiles}
\end{figure*}
Figure \ref{fig:theta-profiles} shows the time evolution of the azimuthally averaged quantities at the stellar surface and in the jet for two different magnetic inclinations. Blue curves show $\chi_{\rm star} = 0^{\circ}$ and red curves show $\chi_{\rm star} = 90^{\circ}$. The dotted and solid lines in panel (a) show $\langle {B^r}^2\rangle$ close to the surface at initial ($t = 0$) and later times ($t = 20000r_g/c$) respectively. Due to the frozen in magnetic boundary conditions, the stellar flux $\langle {B^r}^2\rangle$ at the surface stays nearly constant throughout the simulation. Panel (b) and (c) shows $\langle {B^r}^2\rangle$ (solid lines) and $\langle {T^r}_{t, EM}\rangle$ (dashed lines) in the wind and the jet at an intermediate time (t = $200 r_g/c$)  and after reaching a steady state (t = $20000r_g/c$). At t = $200 r_g/c$, the profiles follow the isolated wind solution except for the equator. The extended cut in the angular profiles is due to the presence of an equatorial disk. Finally, after disk-induced collimation of the open flux, $\langle {B^r}^2\rangle$ has an almost constant angular profile in the jet.   

\section{Star-disk magnetic interaction}\label{appendix3}
\begin{figure*}
    \centering   
     \includegraphics[trim=0 440 0 0,width=15cm, angle=0]{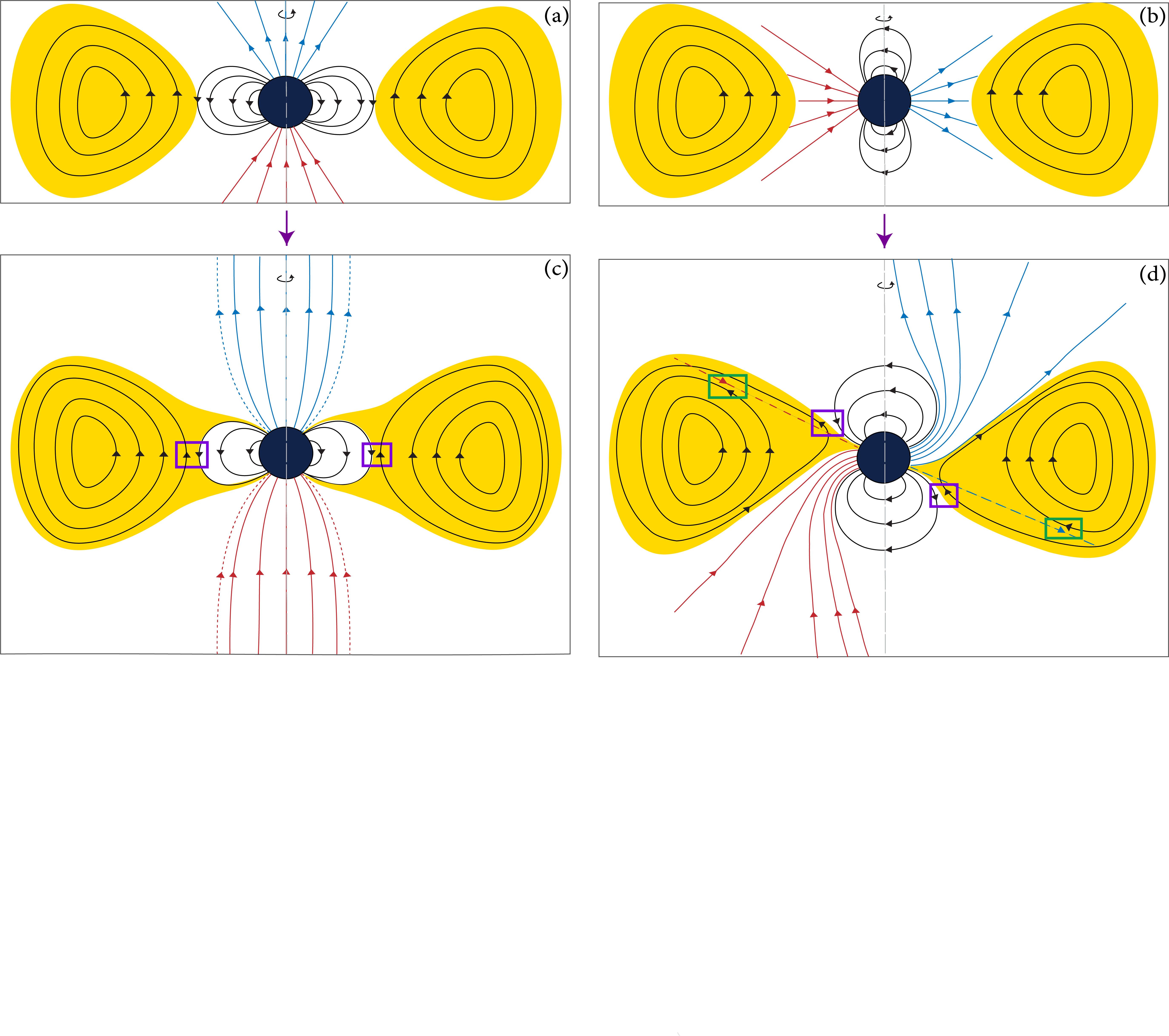}
     \caption{Stellar and disk magnetic field interaction for $\chi_{\rm star} = 0^{\circ}$ (left) and $\chi_{\rm star} = 90^{\circ}$ (right). The upper panel shows the initial magnetic field and the lower panel shows the interaction at an intermediate time. The blue and red lines in the lower panel show the fieldlines in the upper and lower jet respectively. The magnetic reconnection of fieldlines with opposing polarity is indicated by the purple and green boxes. The purple box represents the reconnection of the initially closed stellar zone, while the green box represents reconnection of the initially open stellar fieldlines. The dotted lines in panel (c) represents the newly open fieldlines as a result of reconnection of the stellar closed zone. The dashed lines in panel (d) show how the initially open fieldlines in the lower right and upper left get re-arranged by the disk due to magnetic reconnection. The accretion column has a shifted position as a consequence of the re-arrangement of polar field lines.}
     \label{fig:cartoon}
\end{figure*}
Figure \ref{fig:cartoon} sketches the temporal evolution of the magnetic interaction between the star and the disk. The upper panel (a,b) show the magnetic field configuration at the initial state and the lower panel show the flux rearrangement due to the accretion disk. The color code of the open flux is the same as mentioned in Figure \ref{fig:trac-surfacemap}. In the aligned case ($\chi_{\rm star} = 0^{\circ}$, left panels), the opposite polarity magnetic loops in the disk gets reconnected with the initially closed stellar flux leading to enhanced flux in the jet. The newly opened fieldlines are shown with the dotted lines. In the orthogonal case ($\chi_{\rm star} = 90^{\circ}$, right panels), while some of the initially open stellar flux (dashed lines in panel d) get pushed into the jet, most of them gets absorbed by the disk. Open fieldlines from the right pole gets pushed into the upper jet and ones from the left pole gets pushed into the lower jet. Disk induced opening of the stellar closed zone via magnetic reconnection leads to additional open flux (blue ones from the upper closed zone and red ones from the lower closed zone). 



\end{document}